\documentclass[prl,twocolumn,superscriptaddress,nobibnotes,a4paper]{revtex4-2}
\usepackage[pdftex]{graphicx}
\usepackage[pdftex]{epsfig}
\usepackage{amsmath}
\usepackage{amssymb}
\usepackage{amsfonts}
\usepackage{color}
\usepackage{wrapfig}
\usepackage{eucal}
\usepackage{hhline}
\usepackage{threeparttable}
\usepackage{supertabular}
\usepackage{multirow}
\usepackage{tabularx}
\usepackage[separate-uncertainty=true]{siunitx}
\usepackage{upgreek}
\usepackage{float}
\usepackage{siunitx}

\DeclareMathOperator{\sign}{sign}

\newcommand*\diff{\mathop{}\!\mathrm{d}}
\usepackage{mathtools}

\usepackage[export]{adjustbox}

\usepackage{soul}

\usepackage[version=4]{mhchem} 
\usepackage{array} 
\usepackage[colorlinks=true, allcolors=blue]{hyperref} 

\usepackage{comment}

\setstcolor{red}

\begin{document}

\title{Phononic Crystals in Superfluid Thin-Film Helium}

\author{Alexander Rolf Korsch}
\affiliation{Department of Physics, Fudan University, Shanghai 200433, P.R.\ China}
\affiliation{Department of Physics, School of Science, Westlake University, Hangzhou 310030, P. R.\ China}
\affiliation{Kavli Institute of Nanoscience, Department of Quantum Nanoscience, Delft University of Technology, 2628CJ Delft, The Netherlands}
\author{Niccol\`{o} Fiaschi}
\affiliation{Kavli Institute of Nanoscience, Department of Quantum Nanoscience, Delft University of Technology, 2628CJ Delft, The Netherlands}
\author{Simon Gr\"oblacher}
\email{s.groeblacher@tudelft.nl}
\affiliation{Kavli Institute of Nanoscience, Department of Quantum Nanoscience, Delft University of Technology, 2628CJ Delft, The Netherlands}


\begin{abstract}
In recent years, nanomechanical oscillators in thin films of superfluid helium have attracted attention in the field of optomechanics due to their exceptionally low mechanical dissipation and optical scattering. Mechanical excitations in superfluid thin films -- so-called third sound waves -- can interact with the optical mode of an optical microresonator by modulation of its effective refractive index enabling optomechanical coupling. Strong confinement of third sound modes enhances their intrinsic mechanical non-linearity paving the way for strong phonon-phonon interactions with applications in quantum optomechanics. Here, we realize a phononic crystal cavity confining third sound modes in a superfluid helium film to length scales close to the third sound wavelength. A few nanometer thick superfluid film is self-assembled on top of a silicon nanobeam optical resonator. The periodic patterning of the silicon material creates a periodic modulation of the superfluid film leading to the formation of a phononic band gap. By engineering the geometry of the silicon nanobeam, the phononic band gap allows the confinement of a localized phononic mode.
\end{abstract}

\newpage
\maketitle

\section*{Introduction}

Phononic crystals are structures in which periodic modulation of a material enables the engineering of the propagation of phonons~\cite{m_elastic_1992, kushwaha_acoustic_1993}. Because of their periodic structure, phononic crystals can slow down the propagation speed of phonons or even form acoustic band gaps -- frequency bands in which propagation of phonons is forbidden. Combined with advanced micro- and nanofabrication, phononic crystals enable the realization of low-loss phononic waveguides and high-quality cavities with applications in cavity optomechanics~\cite{eichenfield_optomechanical_2009,aspelmeyer_cavity_2014} and for realizing extremely low-loss mechanical oscillators~\cite{maccabe_nano-acoustic_2020}. Controlling the propagation of phonons furthermore opens the way for engineering the thermal properties of materials~\cite{nomura_review_2022}. Phononic crystals have been realized in a variety of materials ranging from silicon and diamond~\cite{kuruma_engineering_2023}, to graphene~\cite{kirchhof_tunable_2021}, among many others.

In recent years, superfluid helium has attracted attention as a promising material for realizing high-quality mechanical oscillators owing to the absence of viscous damping~\cite{agarwal_theory_2014}. Using acoustic phonons in bulk superfluid $^4$He, mechanical quality factors exceeding $10^7$ have been achieved~\cite{lorenzo_superfluid_2014}. Acoustic phonons in superfluid helium have been utilized in cavity optomechanics experiments to realize Brillouin scattering~\cite{kashkanova_superfluid_2017}, observe quantum fluctuations of the acoustic mode~\cite{shkarin_quantum_2019}, and measure higher-order phonon correlations~\cite{patil_measuring_2022}. In these experiments a microscopic optical cavity is filled with a bulk amount of superfluid helium. The modulation of the refractive index of the superfluid inside the cavity due to density variations caused by the acoustic mode leads to optomechanical coupling. Mechanical modes in bulk superfluid helium have promising applications in the search for dark matter~\cite{baker2023optomechanical,hirschel2023helios}.

An alternative approach is superfluid thin-film optomechanics~\cite{baker_theoretical_2016}:\ a thin-layer (typical thickness $1$ to $\qty{10}{nm}$) of superfluid helium self-assembles on the surface of an optical whispering-gallery-mode microresonator due to van-der-Waals forces. The film supports mechanical excitations in the form of surface waves in which the thickness of the helium film is modulated, so-called 'third sound'~\cite{atkins_third_1959}. The variations of the superfluid film thickness lead to a modulation of the effective refractive index of the optical cavity mode resulting in optomechanical coupling. Using this system, laser control of third sound excitations as well as superfluid Brillouin lasing have been demonstrated~\cite{harris_laser_2016,he_strong_2020}. Furthermore, the optomechanical coupling has been utilized to study quantized vortex dynamics in the superfluid film on top of the optical resonator~\cite{sachkou_coherent_2019}.

The restoring force of third sound excitations in the superfluid film is the van-der-Waals force between the superfluid and the solid surface $F_\mathrm{vdW}=\alpha_\mathrm{vdW}/d_\mathrm{He}^4$, where $\alpha_\mathrm{vdW}$ is the Hamacker constant characteristic of the material~\cite{enss_low-temperature_2005}. The van-der-Waals force depends non-linearly on the helium film thickness $d_\mathrm{He}$ providing an intrinsic mechanical non-linearity of third sound modes. It has been predicted theoretically that strong mechanical non-linearity reaching the single-phonon level can be achieved by strongly confining the third sound mode~\cite{sfendla_extreme_2021}. Single-phonon mechanical non-linearities would allow the preparation of non-classical mechanical states~\cite{rips_steady-state_2012,rips_nonlinear_2014} as well as realizing mechanical qubits for quantum computation~\cite{rips_quantum_2013}. Strong confinement of the third sound mode requires reducing the size of the optical resonator the superfluid film is deposited on. To achieve confinement to length scales on the order of the wavelength, phononic crystals in superfluid thin-films have been proposed~\cite{sfendla_extreme_2021}:~by periodically patterning the substrate material, a periodic modulation of the superfluid thin film is induced, which leads to the formation of phononic band gaps that can be used to confine a phononic third sound mode to length scales around $\qty{100}{nm}$.

Here, we design and experimentally realize a phononic crystal cavity confining third sound modes in superfluid helium thin films on silicon nanobeam optical resonators to length scales close to the third sound wavelength of $\lambda \approx \qty{1}{\micro\meter}$. Phononic cavity modes couple optomechanically to the high-quality optical mode enabling homodyne measurement of the mechanical mode spectrum. We demonstrate control over the mechanical resonance frequency by varying the thickness of the helium film. Lastly, we demonstrate photothermal backaction of the optical mode on the mechanical mode resulting in linewidth broadening and narrowing depending on laser frequency.

\section*{Methods}

\begin{figure}
	\centering
	\includegraphics[width = 8.8cm]{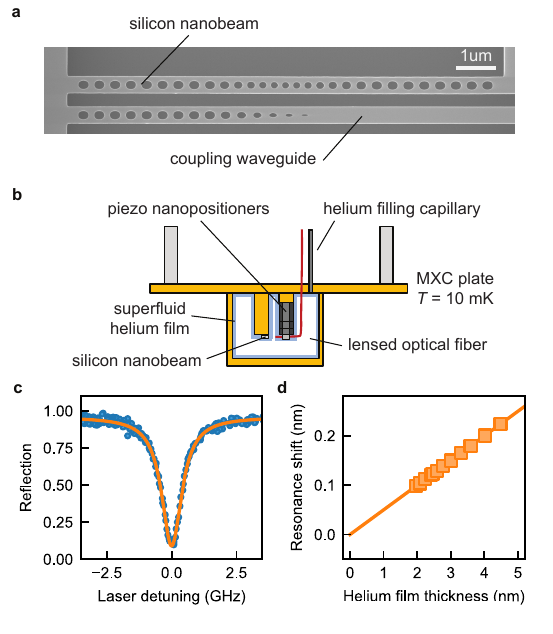}
	\caption{\textbf{Experimental setup.} \textbf{(a)} Scanning electron microscope image of a silicon nanobeam device including coupling waveguide. \textbf{(b)} Schematic illustration of the cryogenic vacuum chamber setup. The chamber made from annealed copper is mounted on the mixing chamber plate of a dilution refrigerator with base temperature $T=10$ mK and contains the silicon nanobeam sample. Light is coupled into and out of the device through a lensed optical fiber. The chamber can be filled with helium gas through a thin stainless steel capillary. \textbf{(c)} Measured reflection spectrum of the optical cavity resonance without superfluid helium when scanning the laser frequency. Solid line corresponds to a Lorentzian fit with center wavelength $\lambda_\mathrm{c} = \qty{1551.10}{nm}$ and linewidth $\kappa=\qty{900}{MHz}$. \textbf{(d)} Shift of the optical resonance wavelength as a function of helium film thickness calibrated using finite-element simulations.}
	\label{Fig:1_experimental_setup}
\end{figure}

Silicon nanobeam resonators are fabricated from a silicon-on-insulator wafer with a device layer thickness of $\qty{250}{nm}$. Using electron beam lithography and HBr-Ar reactive ion etching, structures are patterned in the device layer. The devices are undercut using concentrated hydrofluoric acid to remove the buried oxide layer. The released devices are coated with a $\qty{3}{nm}$ alumina layer using $50$ cycles of atomic layer deposition at a temperature of $\qty{105}{\degreeCelsius}$ preventing further oxidation of the silicon surface. A scanning electron microscope image of a fabricated device is shown in Fig.~\ref{Fig:1_experimental_setup}(a). The nanobeam resonator is evanescently coupled to a center waveguide through which light is coupled to the device.

We embed the silicon nanobeam optical resonators in a cryogenic chamber setup mounted inside a dilution refrigerator at a temperature of $T=\qty{10}{mK}$ as shown in Fig.~\ref{Fig:1_experimental_setup}(b). Laser light is coupled into the coupling waveguide of the device via a lensed optical fiber. Ultra-pure $^4$He gas (99.9999\% purity) filled into the chamber thermalizes to the cryogenic environment and becomes superfluid. As a result of attractive van-der-Waals forces, the superfluid helium covers all surfaces inside the chamber with a thin film of few-nm thickness, including the silicon nanobeam resonator.
 
The silicon nanobeam resonator confines an optical mode at a design wavelength $\lambda=\qty{1550}{nm}$. Figure~\ref{Fig:1_experimental_setup}(c) shows the reflection spectrum of the optical cavity resonance of the silicon nanobeam without superfluid helium. The superfluid helium film changes the effective refractive index of the optical cavity mode causing a red-shift of the resonance frequency. By comparing the measured shift of the optical resonance frequency to results of finite-element simulations, we calibrate the thickness of the superfluid helium film as more helium gas is added into the chamber, as shown in Fig.~\ref{Fig:1_experimental_setup}(d) (see Supplementary Information).

\begin{figure*}
	\centering
	\includegraphics[width = 18cm]{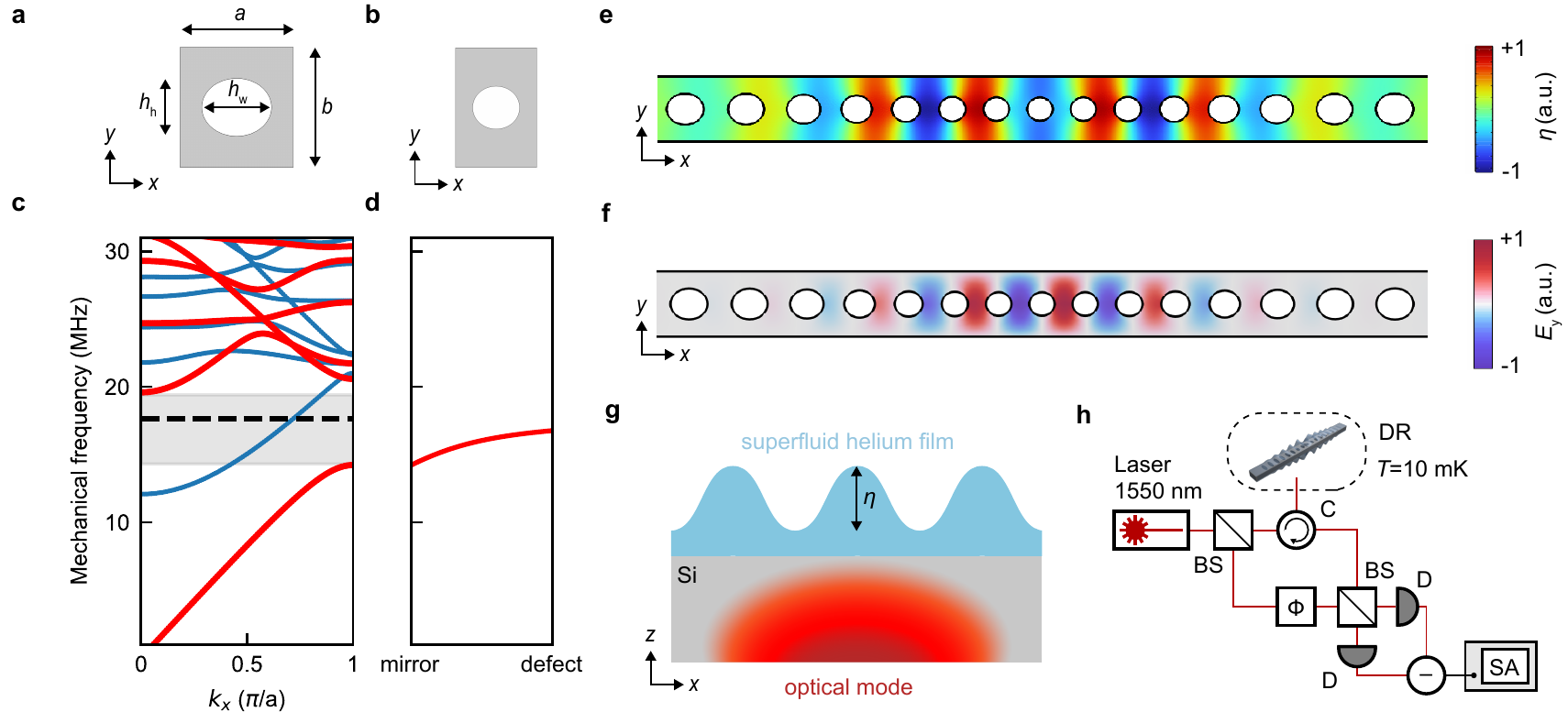}
	\caption{\textbf{Phononic crystals in superfluid thin-film helium.} \textbf{(a)} Phononic crystal unit cell with ($a$, $b$, $h_\mathrm{h}$, $h_\mathrm{w}$)=(472, 499, 241, 290) nm and \textbf{(b)} defect unit cell with ($a$, $b$, $h_\mathrm{h}$, $h_\mathrm{w}$)=(332, 499, 179, 193) nm in superfluid helium thin-film. \textbf{(c)} Third sound band structure of the mirror unit cell for propagation in $x$-direction in a helium film of thickness $d_\mathrm{He}=2.1$ nm. Symmetric (anti-symmetric) modes with respect to the $y$-direction are shown in red (blue). The phononic crystal exhibits a quasi-band gap for symmetric modes for frequencies between $14$ to $19$ MHz (grey shaded). \textbf{(d)} Shift of the bottom band edge at the $X$-point when transforming the mirror unit cell in \textbf{(a)} to a defect unit cell. \textbf{(e)} Normalized variation $\eta$ of the superfluid film thickness in the mechanical mode indicated by the dashed line in \textbf{(c)}. \textbf{(f)} Normalized $y$-component of the electric field $E_y$ in the optical cavity mode. \textbf{(g)} Schematic illustration of optomechanical coupling between the optical mode localized in the silicon nanobeam and a superfluid helium film with thickness variation $\eta$. \textbf{(h)} Optical setup for homodyne detection of third sound modes on the nanobeam. BS, beam splitter; C, circulator; D, photodiode; SA, spectrum analyzer; DR, dilution refrigerator.}
	\label{Fig:2_concept}
\end{figure*}

The periodic patterning of the nanobeam leads to a periodic modulation of the helium film on its surface, which gives rise to a phononic band structure. We model the phononic properties of a periodically modulated superfluid film using finite-element simulations of third sound based on the approach developed in~\cite{forstner_modelling_2019}. Figures~\ref{Fig:2_concept}(a) and (b) show the unit cells of the helium film on the nanobeam surface for the mirror (a) and (b) center defect region. Figure~\ref{Fig:2_concept}(c) shows the resulting band structure of symmetric (red) and anti-symmetric (blue) third sound modes in a helium film of thickness $d_\mathrm{He}=\qty{2.1}{nm}$ for propagation in $x$-direction. In the following, we focus the discussion on $x$-symmetric modes only, as anti-symmetric modes do not couple to the symmetric optical mode due to a vanishing mode overlap. The band structure exhibits a quasi-band gap for symmetric modes between $14$ and $\qty{19}{MHz}$. When changing the parameters of the unit cell from the mirror geometry in (a) to the defect geometry in (b) (see Fig.~\ref{Fig:2_concept}(d)), the bottom edge of the quasi-band gap is continuously shifted into the band gap confining the phononic mode in the defect region. The mode profile of the confined third sound mode is shown in Fig.~\ref{Fig:2_concept}(e).

Figure~\ref{Fig:2_concept}(f) shows the distribution of the electromagnetic field in the fundamental optical mode of the silicon nanobeam. Third sound modes with surface wave amplitude $\eta$ on the nanobeam surfaces modulate the refractive index around the optical cavity mode as schematically illustrated in Fig.~\ref{Fig:2_concept}(g) and therefore lead to optomechanical coupling. The optomechanical interaction transduces thermal motion of third sound modes onto the optical phase quadrature, which we read out using balanced homodyne detection (see Fig.~\ref{Fig:2_concept}(h) for a simplified sketch of the measurement setup).

\section*{Results}

\begin{figure*}
	\centering
	\includegraphics[width = 18cm]{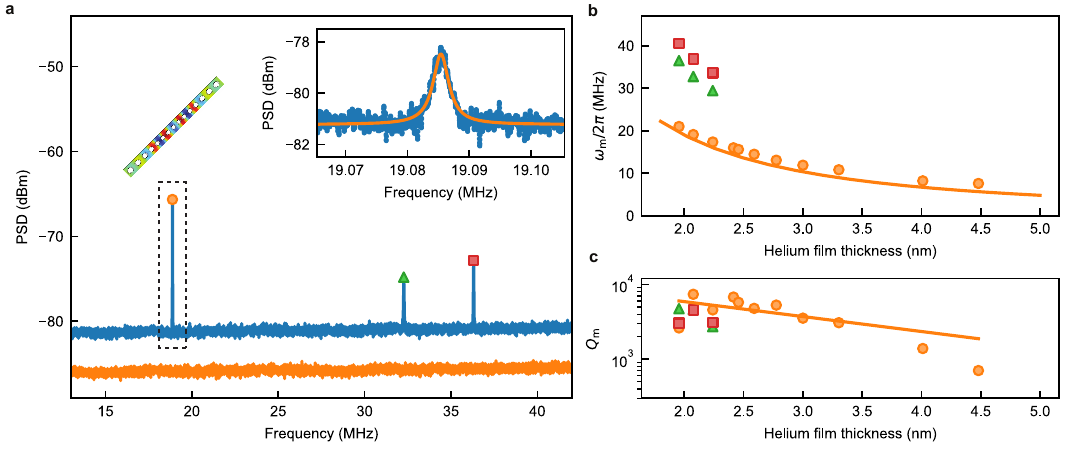}
	\caption{\textbf{Characterization of phononic crystal cavity modes in superfluid helium thin-films.} \textbf{(a)} Noise power spectral density (PSD) of third sound modes in a superfluid helium film of thickness $d_\mathrm{He}=\qty{2.1}{nm}$ (blue) at $\qty{1.6}{nW}$ injected laser power. The spectrum shows three distinct peaks associated to different confined third sound modes. The inset shows a zoom-in of the fundamental phononic mode indicated by the dashed box at frequency $\omega_\mathrm{m}/2\pi=19.085$~MHz. For reference, the spectrum recorded with $\qty{20}{nW}$ laser power before depositing the superfluid helium film is shown (orange). The reference spectrum is offset by $\qty{-5}{dBm}$ for clarity. \textbf{(b)} Mechanical frequency and \textbf{(c)} quality factor of three confined third sound modes as a function of helium film thickness. Solid line in \textbf{(b)} corresponds to results of finite-element simulations, in \textbf{(c)} to a fit using $Q_\mathrm{m} = a \exp \left( -d_\mathrm{He}/l \right)$ with $a=(1.5 \pm 0.8)  \cdot 10^4$ and $l=\qty{2.2 \pm 1.0}{nm}$. Error bars for the data in \textbf{(b)} and \textbf{(c)} are not visible as they are smaller than the data point size.}
	\label{Fig:3_characterization}
\end{figure*}

We measure the mechanical spectrum of helium third sound modes by locking the laser on the optical cavity resonance and using the homodyne detection setup to detect any phase modulation of the light field. Before the superfluid film is deposited on the nanobeam resonator, no mechanical modes are visible in the specturm between $15-40$~MHz (cf,\ Fig.~\ref{Fig:3_characterization}(a), orange line). For a helium film with a thickness of $d_\mathrm{He}=\qty{2.1}{nm}$, the measured mechanical spectrum (blue line) shows three distinct peaks, which we can associate with third sound modes of the helium thin film:\ the lowest frequency mode is the fundamental mechanical mode of the phononic crystal cavity shown in Fig.~\ref{Fig:2_concept}(e). The two peaks at higher frequencies are related to higher-order confined phononic modes of the superfluid film on the nanobeam geometry. However, due to the complex band structure at higher mechanical frequencies (see Fig.~\ref{Fig:2_concept}(a)) combined with fabrication-induced imperfections of the geometry, these two peaks cannot be unambiguously  identified with a specific mode shape from FEM simulations. We show several of these modes and their simulated frequencies in the Supplementary Information. In addition, higher-order modes in the mechanical spectrum can only be observed at the lowest helium film thicknesses due to instability of the helium film caused by driving of extended third sound modes (see Supplementary Information) observed predominantly at higher film thickness. We note that the superfluid phononic crystal modes on the top and bottom surface of the silicon nanobeam remain fully degenerate in frequency within the mechanical linewidth.

We measure the mechanical spectrum as a function of helium film thickness by subsequently adding more helium gas into the chamber. As shown in Fig.~\ref{Fig:3_characterization}(b), the mode frequency of all observed modes in the superfluid film decreases rapidly with increasing film thickness. The restoring force for third sound modes is the van-der-Waals force between the superfluid film and the substrate, which decreases with increasing film thickness as $F_\mathrm{vdW}=\alpha_\mathrm{vdW}/d_\mathrm{He}^4$, where $\alpha_\mathrm{vdW}$ is the Hamacker constant characteristic of the substrate~\cite{enss_low-temperature_2005}. For thicker superfluid helium films, the decrease leads to spring softening and thus to a decrease in mechanical frequency. We perform finite-element simulations of the fundamental third sound mode varying the superfluid film thickness. We fit the model to the measured data with only the Hamacker constant as a free parameter and find good agreement for $\alpha_\mathrm{vdW}=\qty{0.9}{m^5 s^{-2}}$. To the best of our knowledge, this is the first time the Hamacker constant between aluminum oxide and superfluid helium has been measured. Our approach allows to obtain the constant for several different materials in a relatively straight forward way. We note that there is a thin ($\qty{0.5}{nm}$) layer of nitrogen ice covering the surface of the nanobeam due to imperfect vacuum during cooldown which might impact the exact numerical value of the Hamacker constant obtained here.

Figure~\ref{Fig:3_characterization}(c) shows the mechanical quality factor of both the fundamental phononic crystal cavity mode as well as the higher-order modes in the mechanical spectrum. The highest quality factor $Q_\mathrm{m}=7,400$ is observed for the fundamental phononic mode at a film thickness $d_\mathrm{He}=\qty{2.1}{nm}$, while the quality factors of the higher-order modes are comparable to that of the fundamental mode. For increasing film thickness, the mechanical quality factor decreases strongly.

\begin{figure}
	\centering
	\includegraphics[width = 8.8cm]{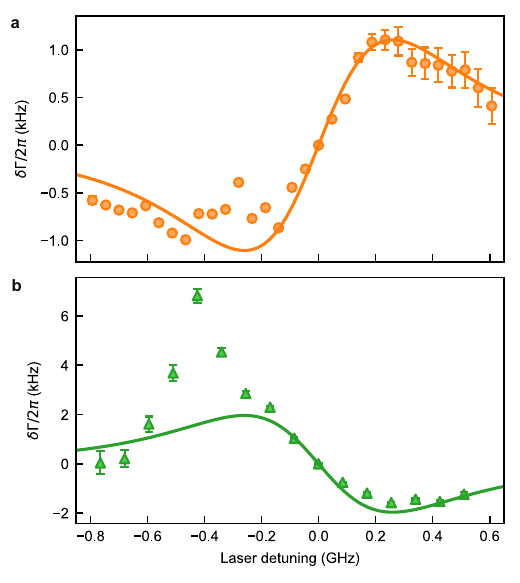}
	\caption{\textbf{Photothermal backaction on helium phononic crystal cavity modes.} Shift of the mechanical linewidth $\delta \Gamma$ as a function of laser detuning for \textbf{(a)} the fundamental, and \textbf{(b)} the first higher-order mechanical mode. When the laser is on resonance with the optical cavity, these modes have mechanical linewidths of $\Gamma_\mathrm{m}=\qty{2.7}{kHz}$ and $\Gamma_\mathrm{m}=\qty{8.2}{kHz}$, respectively. Solid lines correspond to fits obtained from photothermal theory (see Supplementary Information). The sign of the photothermal coupling constant $\sigma$ is negative (positive) for the fundamental (first higher order) mode.}
	\label{Fig:4_photothermal_coupling}
\end{figure}

We now characterize the coupling between the optical and helium third sound modes by measuring the modification of the mechanical linewidth through dynamical backaction. Figure~\ref{Fig:4_photothermal_coupling} shows the change in mechanical linewidth $\delta \Gamma$ for (a) the fundamental phononic crystal cavity mode and (b) the first higher-order mode in the mechanical spectrum as a function of laser detuning from the optical cavity resonance. Remarkably, for the fundamental phononic mode in (a) the mechanical linewidth decreases (increases) when the laser is red-detuned (blue-detuned) from the optical cavity resonance, whereas the higher-order phononic mode exhibits the opposite dependence. Such opposing signs of dynamical backaction between different mechanical modes cannot be explained through only radiation pressure coupling. Similar observations were made for microlevers in atomic force microscopy~\cite{jourdan2008mechanical}, as well as for third sound modes on the surface of whispering-gallery-mode resonators~\cite{harris_laser_2016} and explained in the context of photothermal coupling theory:\ the optical mode creates a hot spot in the center of the silicon nanobeam, which induces flow of the superfluid through the superfluid fountain effect~\cite{allen_new_1938}. The sign of the photothermal coupling constant $\sigma$ depends on the overlap between the superfluid flow field and the respective third sound mode and can thus vary between modes. As derived in the Supplementary Information, photothermal coupling leads to dynamical backaction which modifies the mechanical linewidth. For opposing signs of the photothermal coupling constant, the dependence of mechanical linewidth on laser detuning can be fit using an analytical model, as shown by solid lines in Fig.~\ref{Fig:4_photothermal_coupling}. We find coupling constants of $G/2\pi=\qty{120}{kHz}$ and $G/2\pi=\qty{160}{kHz}$ for the fundamental and first higher-order phononic modes, respectively. Note that the coupling constant given here includes both photothermal coupling as well as dispersive optomechanical interaction due to radiation pressure. The thermalization time constant governing the photothermal interaction of $\tau=\qty{300}{ns}$ obtained from the fit is consistent with previous results in literature~\cite{harris_laser_2016,guenault_detecting_2020}. We attribute the large mechanical linewidth of the higher-order mode in Fig. \ref{Fig:4_photothermal_coupling}(b) around a laser detuning of $\qty{-0.5}{GHz}$ to instability of the helium film resulting from strong driving of extended low-frequency third sound modes.

\section*{Conclusion}

In conclusion, we have designed and experimentally demonstrated a phononic crystal cavity for third sound modes in a superfluid helium thin-film on a silicon nanobeam optical resonator. Confined phononic modes with mechanical quality factors up to $Q_\mathrm{m}=7,400$ were observed through their optomechanical interaction with the optical mode of the silicon nanobeam. The mechanical mode frequency decreases with increasing helium film thickness in good agreement with finite-element simulations. The optomechanical interaction causes dynamical backaction on the mechanical mode which we find is dominated by photothermal coupling.

Previous experiments in superfluid thin-film optomechanics were based on superfluid films assembled on whispering-gallery-mode optical microresonators with diameters on the order of $\qty{50}{\micro\meter}$ to $\qty{100}{\micro\meter}$~\cite{harris_laser_2016,sachkou_coherent_2019,he_strong_2020}. The phononic crystal cavity presented in this work reduces this confinement length by almost two orders of magnitude to about $\qty{5}{\micro\meter}$ for the fundamental phononic crystal cavity mode and $\qty{1}{\micro\meter}$ for select higher-order modes (see Supplementary Information), bringing superfluid third sound modes close to the regime of strong mechanical non-linearity ,where the single-phonon nonlinear frequency shift of the mechanical mode $\delta \Omega_\mathrm{m}$ is larger than the mechanical linewidth $\Gamma_\mathrm{m}$. Based on finite-element simulations, we estimate $\delta \Omega_\mathrm{m}/2\pi = \qty{7}{mHz}$ for the fundamental phononic crystal cavity mode and $\delta \Omega_\mathrm{m}/2\pi = \qty{88}{mHz}$ for the higher order mode with shortest confinement length (see Supplementary Information). Reaching the single-phonon non-linear regime requires both stronger confinement of the phononic mode to an even smaller mode volume as well, as an increased mechanical quality factor. A modest increase of the mechanical confinement to defect sizes on the order of $\qty{100}{nm}$ and an increased mechanical quality factor of around $Q_\mathrm{m}=10^6$ will enable non-linearities on the single-phonon level~\cite{sfendla_extreme_2021}. Both should be within reach by designing phononic crystal structures which are optimized for third sound modes and not for optomoechanical operation. In particular, reducing phonon radiation loss through the phononic mirrors by surrounding the nanobeam with a phononic shield with full phononic band gap similar to phononic shield structures established in silicon optomechanical crystals~\cite{chan_optimized_2012,maccabe_nano-acoustic_2020, wallucks_quantum_2020} will significantly boost the mechanical quality factor. Bringing the system close to the single-phonon non-linear regime will first manifest in asymmetric mechanical lineshapes and eventually in the splitting of the mechanical spectrum~\cite{sfendla_extreme_2021}. Finally reaching the single-phonon non-linear regime will open up new approaches for creating non-classical states of mechanical motion~\cite{rips_steady-state_2012,rips_nonlinear_2014} and pave the way for realizing mechanical qubits for quantum computation~\cite{rips_quantum_2013}.

\medskip

\textbf{Acknowledgments}
We would like to thank Xiong Yao and Liu Chen for helpful discussions. We further acknowledge assistance from the Kavli Nanolab Delft. This work is financially supported by the European Research Council (ERC CoG Q-ECHOS, 101001005) and is part of the research program of the Netherlands Organization for Scientific Research (NWO), supported by the NWO Frontiers of Nanoscience program, as well as through a Vrij Programma (680-92-18-04) grant.

\textbf{Conflict of interests:}\ The authors declare no competing interests.

\textbf{Author contributions:}\  A.R.K., N.F., and S.G., devised and planned the experiment. A.R.K., and N.F. built the setup and performed the measurements. A.R.K. fabricated the device. All authors analyzed the data and wrote the manuscript. S.G. supervised the project.

\textbf{Data Availability:}\ Source data for the plots are available on \href{Zenodo}{Zenodo}.

\setcounter{figure}{0}
\renewcommand{\thefigure}{S\arabic{figure}}
\setcounter{equation}{0}
\renewcommand{\theequation}{S\arabic{equation}}

\clearpage
\newpage

\onecolumngrid

\begin{center}
	\textsc{\Large Supplementary Information} 
\end{center}
\label{SI}

\subsection{Details of the cryogenic chamber setup}
\label{experimental_setup_details}

Experiments are performed in a cryogenic chamber setup mounted on the bottom of the mixing chamber plate of a dilution refrigerator, reaching a base temperature of $T=10$~mK. The chamber is made of annealed copper and sealed using indium wire seals. The bottom lid of the chamber is made of regular copper to avoid bending due to the difference in pressure inside and outside the chamber. Electrical feedthroughs for the piezo-nanopositioners are realized using a D-sub connector silver-soldered on a stainless steel plate sealed with indium wire to the copper body of the chamber. A stainless steel capillary is attached to the top of the chamber using silver soldering. The capillary is used for both pumping vacuum in the chamber and to load the helium. To avoid a large thermal load between the stages of the dilution refrigerator, we choose the capillary to have a small diameter ($1/16~\mathrm{inch}$), which, together with its length  of approx.\ 1~m, limits the base pressure in the chamber to around $\qty{0.1}{mbar}$. For this reason we use nitrogen to flush the chamber to remove any air before each cooldown. We pump down and repeat this process $10$ times before starting the cooldown. During cooldown the remaining nitrogen partially freezes on the surface of the silicon chip leaving a thin ($\sim$$\qty{0.5}{nm}$) layer of nitrogen ice.

Lensed optical fibers are inserted into small holes in the top of the chamber and sealed with Stycast epoxy resin. Inside the chamber, the silicon chip containing the nanobeam devices is mounted on a copper sample holder. The tip of the lensed optical fiber is mounted on a stack of piezo nanopositioners which allow to couple light into a waveguide, which in turn is evanescently coupled to the nanobeam devices~\cite{riedinger_non-classical_2016}.

\subsection{Calibration of helium film thickness}
\label{helium_film_thickness_calibration}

The thickness of the film can be controlled by the amount of helium gas inserted into the chamber. Once the thickness of the superfluid film reaches a critical film thickness $d_\mathrm{crit}$, additional inserted helium gas accumulates in the bottom of the chamber. The critical film thickness is set by the equilibrium between gravitational and van-der-Waals forces and depends on the height difference $z_\mathrm{chip}$ between the silicon chip and the bottom of the chamber \cite{enss_low-temperature_2005}:
\begin{align}
	\label{critical_film_thickness}
	d_\mathrm{crit} = \left(\frac{\alpha_\mathrm{vdW}}{g z_\mathrm{chip}}\right)^{1/3},
\end{align}
where $g=\qty{9.81}{m.s^{-2}}$ is the acceleration due to gravity and $\alpha_\mathrm{vdW}=\qty{0.9e-24}{m^5.s^{-2}}$ is the Hamacker constant as identified by our experiments. In our setup, $z_\mathrm{chip} \approx \qty{3}{mm}$, resulting in a critical film thickness $d_\mathrm{crit} = \qty{30}{nm}$. In all experiments performed in this work, we operate in a regime of very thin films far below this critical film thickness.

To calibrate the thickness of the superfluid film, we measure the shift of the optical cavity resonance frequency as more helium gas is added into the chamber. We note that we use low optical laser power of around $P_\mathrm{laser}=\qty{10}{fW}$ and detect the reflected photons on superconducting nanowire single-photon detectors (SPDs) to avoid any effect of the optical pumping on the cavity resonance frequency through interaction with the superfluid helium film as discussed in section~\ref{time_resolved_detection}.

\begin{figure}
	\centering
	\includegraphics[width = 0.5\linewidth]{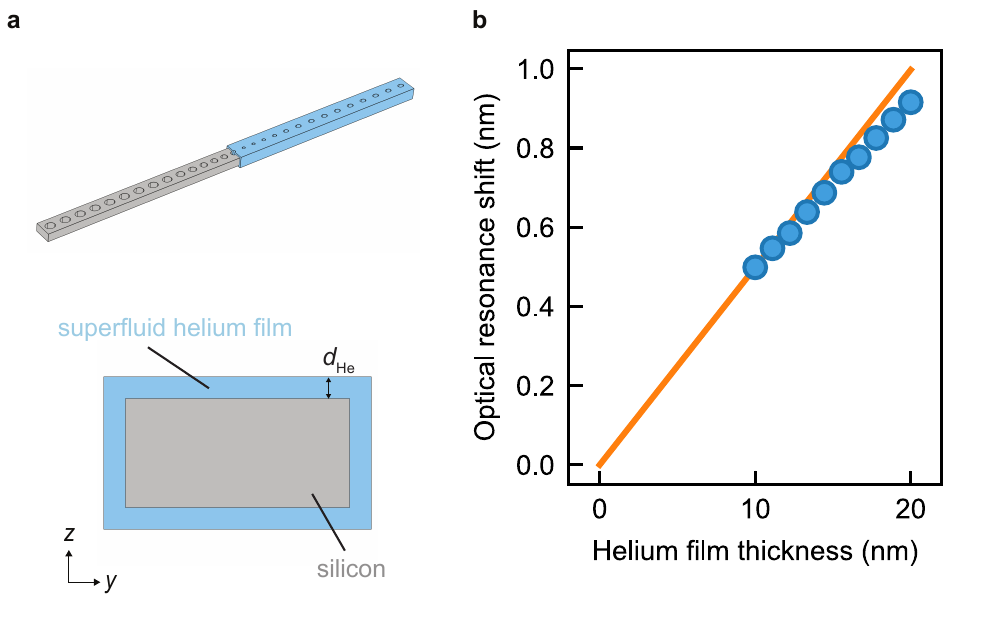}
	\caption{\textbf{Calibration of the superfluid helium film thickness.} \textbf{(a)} Geometry for COMSOL finite-element simulation (top) and schematic cross-section (bottom) of a silicon nanobeam covered in a thin-film of superfluid helium of thickness $d_\mathrm{He}$. \textbf{(b)} Simulated shift of the optical resonance wavelength when increasing the helium film thickness. The orange line is a linear extrapolation of the simulated data points for low film thickness corresponding to a shift of the optical resonance wavelength $\qty{0.05}{nm}$ per nanometer helium film thickness.}
	\label{Fig:S1_helium_film_thickness_calibration}
\end{figure}

We perform finite-element simulations using COMSOL Multiphysics to determine the expected optical resonance shift. Figure \ref{Fig:S1_helium_film_thickness_calibration}(a) shows the geometry used for COMSOL simulations where all surfaces of the silicon nanobeam are covered with a thin superfluid film of thickness $d_\mathrm{He}$ and refractive index $n_\mathrm{He}=1.026$~\cite{edwards_refractive_1958}. Since simulations of very thin films $d_\mathrm{He} < \qty{10}{nm}$ require very fine meshing of the geometry and therefore long simulation times, we simulate the optical resonance shift only for $d_\mathrm{He} \geq \qty{10}{nm}$ and extrapolate the result to the case without a superfluid film. Figure~\ref{Fig:S1_helium_film_thickness_calibration}(b) shows the simulated optical resonance shift as a function of helium film thickness as well as the linear extrapolation to thinner films. This linear extrapolation yields a change of the optical resonance wavelength of $\qty{0.05}{nm}$ per nm helium film thickness. For very thin films we use the extrapolation to obtain the calibrated helium film thickness for each measurement configuration in the main text.

\subsection{Finite-element simulations of third sound modes in superfluid helium}
\label{FEM_simulations_superfluid}

We perform finite-element simulations using the software COMSOL to solve the hydrodynamic equations of third sound modes in the superfluid helium thin-film. Since COMSOL does not provide a dedicated simulation package for superfluid third sound, Forstner et al.\ developed an approach based on the formal analogy of the superfluid hydrodynamic equations to the linearized Euler equation for an ideal gas~\cite{forstner_modelling_2019}. The linearized Euler equation for the superfluid film height $d_\mathrm{He}$ with thickness variations due to third sound $\eta$ and velocity field $\vec{v}$ is given by
\begin{align}
	\label{continuity_equation_third_sound}
	\dot{\vec{v}} + \left(\vec{v} \cdot \vec{\nabla} \right) \vec{v} = -\frac{3 a_\mathrm{vdW}}{d_\mathrm{He}^4} \vec{\nabla}\eta.
\end{align}
The linearized Euler equation for flow of an ideal gas with acoustic velocity field $\vec{u}$, density $\rho_0$, heat capacity ratio $\gamma$, ideal gas constant $R$, temperature $T$, and density perturbation $\alpha = \rho-\rho_0$ is given by
\begin{align}
	\label{linearized_euler_equation_acoustics}
	\dot{\vec{u}} + \left(\vec{u} \cdot \vec{\nabla} \right) \vec{u} = -\frac{\gamma R T}{\rho_0} \vec{\nabla}\alpha.
\end{align}
Equations~\eqref{continuity_equation_third_sound} and~\eqref{linearized_euler_equation_acoustics} are formally analogous to one another. Hence, we can implement simulations of superfluid third sound by solving the linearized Euler equations of an ideal gas substituting $\gamma R T/\rho_0 \rightarrow 3 a_\mathrm{vdW}/d_\mathrm{He}^4$ and $\rho_0 \rightarrow d_\mathrm{He}$. The density perturbation $\alpha$ then corresponds to the variation of the superfluid film height $\eta$. The simulations are implemented using the COMSOL module $\textit{Aeroacoustics} \rightarrow \textit{Linearized Euler, Frequency Domain (lef)}$. Free boundary conditions are set at all boundaries of the superfluid film except the two ends of the nanobeam where fixed boundary conditions are applied. As discussed in \cite{forstner_modelling_2019}, fixed boundary conditions correspond to the \textit{fixed pressure} condition and free boundary conditions correspond to the \textit{rigid wall} condition in the \textit{Linearized Euler} module in COMSOL.

\subsection{Simulations of higher-order third sound modes}
\label{simulations_higher_order_modes}

\begin{figure}
	\centering
	\includegraphics[width = .8\linewidth]{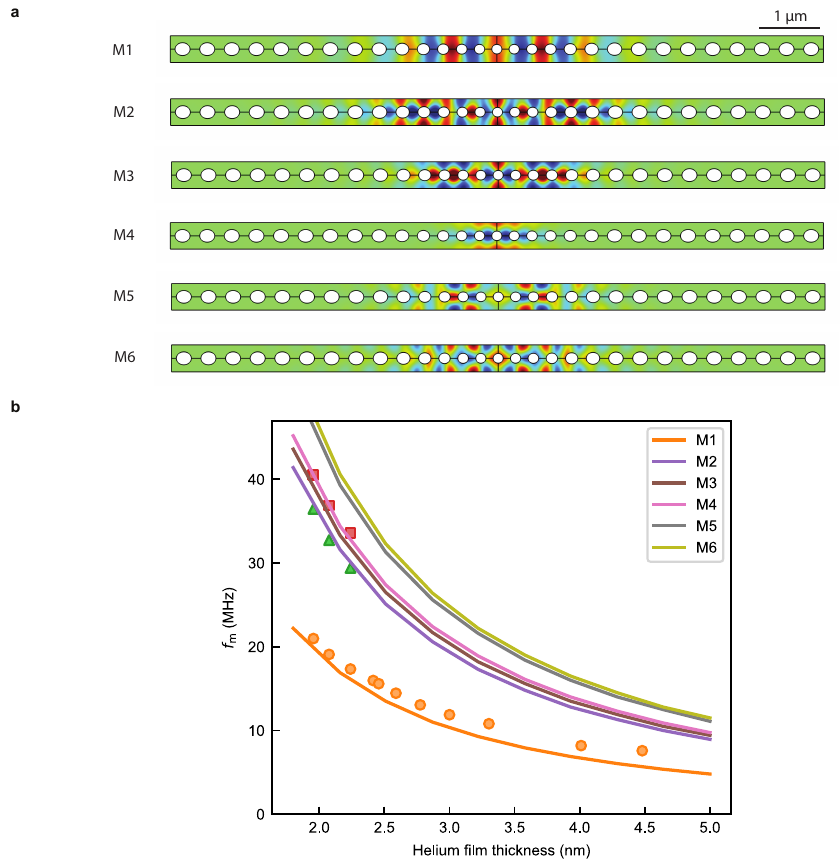}
	\caption{\textbf{FEM simulations of higher-order third sound modes.} \textbf{(a)} Simulated mode profile of thickness variation $\eta$ of different third sound modes M1 to M6. Mode M1 corresponds to the fundamental phononic crystal cavity mode discussed in the main text. \textbf{(b)} Simulated mechanical frequency of third sound modes M1 to M6 as a function of helium film thickness plotted alongside the experimental data shown in the main text.}
	\label{Fig:S7_simulations_higher_order_modes}
\end{figure}

In Figure~\ref{Fig:S7_simulations_higher_order_modes}(a) we show the simulated mechanical mode profiles of six confined third sound modes labeled M1 to M6. Here, the mode labeled M1 corresponds to the fundamental phononic crystal cavity mode discussed in the main text. We also simulate the dependence of the mechanical frequency on helium film thickness for the various modes, as shown in Fig.~\ref{Fig:S7_simulations_higher_order_modes}(b).

\subsection{Calculation of mechanical non-linearity}
\label{calculation_mechanical_nonlinearity}

Sfendla et al.\ developed a theoretical framework for estimating the non-linearity of helium third sound modes originating from the inherently non-linear van-der-Waals interaction with the substrate~\cite{sfendla_extreme_2021}. Here, we use this framework to numerically estimate the mechanical non-linearity of the fundamental phononic crystal cavity mode discussed in the main text. The authors derive the potential energy $U$ stored within the third sound wave as
\begin{align}
	\label{potential_energy}
	U = \rho \int \int \left( \frac{3a_\mathrm{vdW} \eta^2[x,y]}{2d^4} - \frac{2a_\mathrm{vdW} \eta^3[x,y]}{d^5} + \frac{5a_\mathrm{vdW} \eta^4[x,y]}{2d^6} \right) \diff x \diff y.
\end{align}
Here, $\eta[x,y]$ is the thickness variation of the third sound that can be obtained by finite-element simulations, $a_\mathrm{vdW}$ is the Hamacker constant, $d$ is the helium film thickness and $\rho$ is the density of superfluid helium. Note that we choose to use cartesian coordinates $x$, and $y$, instead of cylindrical coordintates used in~\cite{sfendla_extreme_2021}. By introducing a reference point $q \coloneqq \max \left(\eta[x,y]\right)$, Eq.~\eqref{potential_energy} can be written as
\begin{align}
	\label{potential_energy_2}
	U = \frac{1}{2} k q^2 + \frac{1}{3} \beta q^3 + \frac{1}{4} \alpha q^4,
\end{align}
which is the potential energy of an anharmonic oscillator. Here, $k$ corresponds to a linear spring constant, whereas $\beta$ and $\alpha$ quantify the strength of a cubic and quartic non-linearity, respectively. The constants $k$, $\beta$, and $\alpha$ are obtained from integrals of the thickness variation $\eta[x,y]$ and are given by
\begin{align}
	\label{k_beta_alpha}
	k&=\frac{3 \rho a_\mathrm{vdW}}{d^4} \int \int \frac{\eta^2[x,y]}{\max \left(\eta[x,y] \right)^2}, \\
	\beta&=-\frac{6 \rho a_\mathrm{vdW}}{d^5} \int \int \frac{\eta^3[x,y]}{\max \left(\eta[x,y] \right)^3}, \\
	\alpha&=\frac{10 \rho a_\mathrm{vdW}}{d^6} \int \int \frac{\eta^4[x,y]}{\max \left(\eta[x,y] \right)^4}.
\end{align}
To quantify the mechanical non-linearity we consider the induced single-phonon frequency shift $\delta \Omega_\mathrm{m}$ given by
\begin{align}
	\label{single_phonon_frequency_shift}
	\delta \Omega_\mathrm{m} = \frac{3 x_\mathrm{zpf} \alpha_\mathrm{eff}}{\hbar},
\end{align}
with the zero-point motion of the mechanical mode $x_\mathrm{zpf}=\sqrt{\hbar/2m_\mathrm{eff}\Omega_\mathrm{m}}$, effective mass $m_\mathrm{eff}=k/\Omega_\mathrm{m}^2$, and $a_\mathrm{eff}$ defined as
\begin{align}
	\label{a_eff}
	a_\mathrm{eff} = a-\frac{10}{9} \frac{\beta^2}{k}.
\end{align}
It is important to note that the reference point $q \coloneqq \max \left(\eta[x,y]\right)$ is chosen according to the convention usually used for phononic crystal cavities~\cite{eichenfield_optomechanical_2009}. In the original derivation discussed in~\cite{sfendla_extreme_2021}, the authors consider third sound modes confined on a circular geometry of radius $R$ such as whispering-gallery-mode optical microresonators. In these systems, optomechanical coupling occurs at the periphery of the geometry and hence the reference point is chosen to be the thickness variation at the periphery $\eta(R)$. The choice of the reference point impacts the numerical values of the harmonic and anharmonic spring constants $k$, $\alpha$, and $\beta$,  and consequently the zero-point motion $x_\mathrm{zpf}$ and effective mass $m_\mathrm{eff}$. Importantly, the single-phonon frequency shift is independent of the choice of reference point as the respective factors cancel in Eq.~\eqref{single_phonon_frequency_shift}.

\subsection{Theory of photothermal coupling}
\label{photothermal_coupling_theory}

Mechanical motion of the helium modes is actuated mostly through photothermal coupling induced by absorption of photons in the optical cavity of the nanobeam. At the same time, the optical cavity resonance is dispersively coupled to the mechanical motion through modulation of the effective refractive index. To describe the modifications of the mechanical oscillator response through photothermal coupling, we follow the approach outlined in~\cite{harris_laser_2016}. We start from the quantum Langevin equation for a mechanical oscillator with amplitude $x(t)$, effective mass $m_\mathrm{eff}$, frequency $\omega_\mathrm{m}$ and decay rate $\Gamma_\mathrm{m}$. We include a dispersive radiation pressure interaction with the optical cavity field described by annihilation operator $a(t)$ and a photothermal coupling term as well as the thermal Langevin force $F_\mathrm{th}$
\begin{align}
	\label{langevin_mechanics}
	m_\mathrm{eff} [\ddot{x}(t) + \Gamma_\mathrm{m} \dot{x}(t) + \omega_\mathrm{m}^2x(t)] = \hbar g \left[ a(t)^\dag a(t) + \frac{\beta A}{\tau} \int_{-\infty}^{\infty} du \Theta(t) e^{-\frac{t-u}{\tau}} a(u)^\dag a(u)\right] + F_\mathrm{th}(t),
\end{align}
where $\Theta(t)$ is the Heaviside function, $A$ the absorption coefficient of the material, $g=g_0/x_\mathrm{zpf}$ the dispersive optomechanical coupling parameter with mechanical zero-point fluctuations $x_\mathrm{zpf}^2 = \hbar/(2 m_\mathrm{eff} \omega_\mathrm{m})$, and $\beta$ a dimensionless parameter that determines the strength of the photothermal relative to the radiation pressure coupling term.

The optical cavity field is similarly described by the quantum Langevin equation in the frame rotating with $\Delta^0 = \omega_\mathrm{L} - \omega_\mathrm{c}$
\begin{align}
	\label{langevin_optics}
	\dot{a}(t) = -[\kappa/2 - i(\Delta^0 + gx(t))] a(t) + \sqrt{\kappa_\mathrm{ex}} a_\mathrm{in}(t),
\end{align}
where $\kappa$ is the linewidth of the optical cavity, $\kappa_\mathrm{ex}$ the extrinsic cavity coupling and $a_\mathrm{in}(t)$ the input field incident on the optical cavity. We linearize the coupled quantum Langevin equations by considering fluctuations $\delta x(t)$ and $\delta a(t)$ around the equilibrium mechanical oscillator position $\Bar{x}$ and intracavity field $\alpha$, such that $x(t) = \Bar{x}+\delta x(t)$ and $a(t) = \alpha +\delta a(t)$ as well as $a_\mathrm{in}(t) = \alpha_\mathrm{in} +\delta a_\mathrm{in}(t)$. The coupling to the mechanical oscillator leads to a shift of the equilibrium optical cavity resonance
\begin{align}
	\label{alpha}
	\alpha = \frac{\sqrt{\kappa_\mathrm{ex}} \alpha_\mathrm{in}}{\kappa/2 - i \Delta},
\end{align}
where the detuning is given by $\Delta = \Delta^0 + g \Bar{x}$. In the following, we consider the regime where the photothermal coupling is dominant over radiation pressure effects, such that $\beta A/(1+i\omega \tau) \gg 1$. After linearization of Eqs.~\eqref{langevin_mechanics} and \eqref{langevin_optics}, we obtain the solution in the Fourier domain (see~\cite{harris_laser_2016} for details). Fluctuations of the mechanical oscillator are described by
\begin{align}
	\label{delta_x2}
	\delta x = \chi^{\prime}(\omega) F_\mathrm{th}(\omega),
\end{align}
where the modified mechanical susceptibility $\chi^{\prime}(\omega)$ is introduced
\begin{align}
	\label{chi_prime}
	\chi^{\prime}(\omega) &= \frac{m_\mathrm{eff}^{-1}}{m_\mathrm{eff}^{-1}\chi(\omega)^{-1} + K(\omega)}
\end{align}
with the mechanical susceptibility of the uncoupled mechanical oscillator $\chi(\omega)^{-1} = m_\mathrm{eff} (\omega_\mathrm{m}^2 - \omega^2 + i\omega \Gamma_\mathrm{m})$. Modifications to the mechanical oscillator response due to photothermal coupling are contained in $K(\omega)$ which is defined as
\begin{align}
	\label{K}
	K(\omega) &= \frac{4g_0^2 |\alpha|^2 \omega_\mathrm{m} \Delta}{D(\omega) D^*(-\omega)} \frac{\beta A}{1+i\omega \tau},
\end{align}
where we define the transfer function of the optical cavity $D(\omega) = \kappa/2 + i(\omega - \Delta)$. Changes in mechanical frequency $\delta \omega_\mathrm{m}^2$ and decay rate $\delta \Gamma_\mathrm{m}$ are then related to the real and imaginary part of $K(\omega)$
\begin{align}
	\label{Re_Im}
	\delta \omega_\mathrm{m}^2 &= \mathrm{Re}(K(\omega)), \\
	\delta \Gamma_\mathrm{m} &= \mathrm{Im}(K(\omega))/\omega.
\end{align}

\subsubsection{Modifications to the mechanical resonance frequency}
We calculate explicitly
\begin{align}
	\label{delta_omega_square}
	\delta \omega_\mathrm{m}^2 &= \mathrm{Re}(K(\omega)) \\
	&= 4g_0^2 |\alpha|^2 \omega_\mathrm{m} \Delta \beta A \frac{1}{1+\omega^2\tau^2} \frac{\Delta^2 + (\kappa/2)^2 - \omega^2 -\kappa \tau \omega^2}{|D(\omega)D^*(-\omega)|^2}.
\end{align}
To obtain $\delta \omega_\mathrm{m}$, we linearize $\delta \omega_\mathrm{m}^2$ as 
\begin{align}
	\label{delta_omega}
	\delta \omega_\mathrm{m} &\approx \frac{\delta \omega_\mathrm{m}^2}{2\omega_\mathrm{m}} \\
	&= \frac{2g_0^2 |\alpha|^2 \Delta \beta A}{1+\omega^2\tau^2} \frac{\Delta^2 + (\kappa/2)^2 - \omega^2 -\kappa \tau \omega^2}{|D(\omega)D^*(-\omega)|^2}.
\end{align}
As long as the modifications to the mechanical frequency are small, we can evaluate Eq.~\eqref{delta_omega} at $\omega = \omega_\mathrm{m}$
\begin{align}
	\label{delta_omega_final}
	\delta \omega_\mathrm{m} &= \frac{2g_0^2 |\alpha|^2 \Delta \beta A}{1+\omega_\mathrm{m}^2\tau^2} \frac{\Delta^2 + (\kappa/2)^2 - \omega_\mathrm{m}^2 -\kappa \tau \omega_\mathrm{m}^2}{|D(\omega_\mathrm{m})D^*(-\omega_\mathrm{m})|^2}.
\end{align}

\subsubsection{Modifications to the mechanical decay rate}
We calculate explicitly
\begin{align}
	\label{delta_Gamma}
	\delta \Gamma_\mathrm{m} &= \mathrm{Im}(K(\omega))/\omega \\
	&= -\frac{4g_0^2 |\alpha|^2 \omega_\mathrm{m} \Delta \beta A }{1+\omega^2\tau^2} \frac{\kappa + \tau \left[\Delta^2 + (\kappa/2)^2-\omega^2\right]}{|D(\omega)D^*(-\omega)|^2}.
\end{align}
As long as the modifications to the mechanical frequency are small, we can evaluate Eq.~\eqref{delta_Gamma} at $\omega = \omega_\mathrm{m}$
\begin{align}
	\label{delta_Gamma_final}
	\delta \Gamma_\mathrm{m} &= -\frac{4g_0^2 |\alpha|^2 \omega_\mathrm{m} \Delta \beta A }{1+\omega_\mathrm{m}^2\tau^2} \frac{\kappa + \tau \left[\Delta^2 + (\kappa/2)^2-\omega_\mathrm{m}^2\right]}{|D(\omega_\mathrm{m})D^*(-\omega_\mathrm{m})|^2}.
\end{align}

In the experiment, we cannot separate the contributions of the radiation pressure and photothermal interactions to the dynamical backaction effects on the mechanical oscillator response. To simplify the discussion, we combine the coupling parameters $g_0$, $\beta$, and $A$ in a joint coupling parameter $G$ and the sign of the photothermal interaction $\sigma$ such that $\sigma G^2 = g_0^2 \beta A$ with $\sigma=\sign \beta$. We finally obtain the analytical expression
\begin{align}
	\label{joint_coupling_parameter}
	\delta \Gamma_\mathrm{m} &= -\frac{4 \sigma G^2 |\alpha|^2 \omega_\mathrm{m} \Delta}{1+\omega_\mathrm{m}^2\tau^2} \frac{\kappa + \tau \left[\Delta^2 + (\kappa/2)^2-\omega_\mathrm{m}^2\right]}{|D(\omega_\mathrm{m})D^*(-\omega_\mathrm{m})|^2}.
\end{align}
We use this expression to fit the experimental data in the main text using $G$, $\sigma$ and $\tau$ as fitting parameters. The time constant $\tau$ is constrained to be the same for both third sound modes.

\subsection{Shift of the mechanical resonance frequency due to laser-induced heating}
\label{heating_mechanical_frequency}

\begin{figure}
	\centering
	\includegraphics[width = 0.85\linewidth]{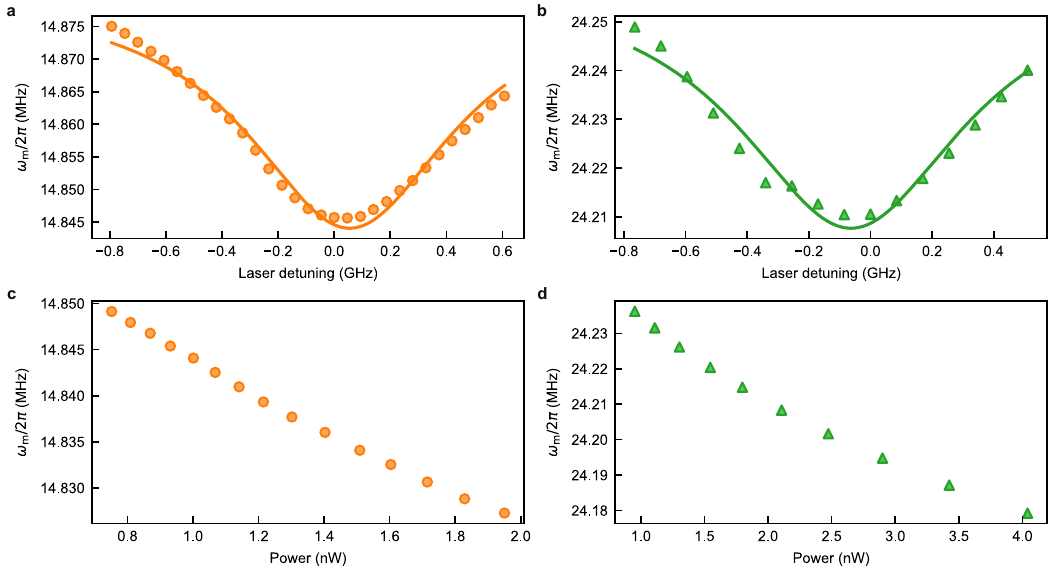}
	\caption{\textbf{Laser-induced heating of helium phononic crystal cavity modes.} \textbf{(a)}, \textbf{(b)} Mechanical resonance frequency $\omega_\mathrm{m}/2\pi$ as a function of laser detuning for \textbf{(a)} the fundamental, and \textbf{(b)} the first higher order mechanical mode. The injected powers into the device are fixed at $\qty{1.5}{nW}$ in \textbf{(a)} and $\qty{1.9}{nW}$ in \textbf{(b)}. Solid lines correspond to Lorentzian fits with a fixed linewidth of $\kappa/2\pi=\qty{900}{MHz}$ corresponding to the optical cavity linewidth. \textbf{(c)}, \textbf{(d)} Mechanical resonance frequency of \textbf{(c)} the fundamental, and \textbf{(d)} the first higher order mechanical mode as a function of injected laser power into the device when the laser is locked on the optical cavity resonance. Error bars for the data are not visible as they are smaller than the data point size.}
	\label{Fig:S2_heating_mechanical_frequency}
\end{figure}

From Eq.~\eqref{delta_omega_final}, we expect a similar effect of dynamical backaction on the mechanical resonance frequency as on the mechanical linewidth discussed in the main text. However, such a dependence is not observed in the experiment:\ Figure~\ref{Fig:S2_heating_mechanical_frequency} shows the mechanical resonance frequency of the fundamental phononic mode (a) and first higher order mode as a function of laser detuning from the optical cavity resonance. For both modes, the mechanical resonance frequency decreases when the laser frequency is tuned closer to the optical cavity resonance. Solid lines in (a) and (b) are Lorentzian fits with the linewidth fixed at the optical resonance linewidth $\kappa/2\pi=\qty{900}{MHz}$, showing good agreement with the experimental data. We further measure the dependence of the mechanical frequency on injected laser power into the device with the laser locked on resonance with the optical cavity mode as shown in Figs.~\ref{Fig:S2_heating_mechanical_frequency}(c),(d). For both mechanical modes, the mechanical frequency decreases with increasing injected laser power.

Similar shifts to lower mechanical resonance frequencies with increasing optical power were observed for superfluid helium films on whispering-gallery-mode optical microresonators and attributed to heating of the superfluid helium film~\cite{harris_laser_2016}:\ when the helium film heats up, the density of the superfluid component decreases, reducing the speed of third sound and thus the mechanical mode frequency. 

While current measurements do not allow a definite conclusion about the source of the heating, it is likely that heating is caused by absorption of photons in the silicon which then transfers heat to the superfluid film, rather than direct absorption in the helium:\ absorption heating in silicon is a well-known effect in silicon optomechanical crystal cavities~\cite{meenehan_silicon_2014,meenehan_pulsed_2015}. Typical time scales for this heating process are in the range of $\qty{100}{ns}$ to $\qty{1}{\micro\second}$~\cite{meenehan_pulsed_2015,fiaschi_optomechanical_2021} in good agreement with the time constant of photothermal coupling $\tau=\qty{300}{ns}$ identified in the main text.

This proposed heating mechanism also explains the observation of decreased mechanical resonance frequency when the laser frequency is tuned closer to the optical resonance in Fig.~\ref{Fig:S2_heating_mechanical_frequency}(a) and (b). In these measurements, the optical power injected into the device $P_\mathrm{in}$ is fixed. Hence, the intracavity photon number inside the optical cavity $n_\mathrm{cav}$ changes as a function of laser detuning $\Delta$ as
\begin{align}
	\label{intracavity_photon_number}
	n_\mathrm{cav} = \frac{\kappa_\mathrm{ex}}{\Delta^2 + (\kappa/2)^2} \frac{P_\mathrm{in}}{\hbar \omega_\mathrm{L}}
\end{align}
with laser frequency $\omega_\mathrm{L}$ and extrinsic cavity coupling $\kappa_\mathrm{ex}=\kappa/2$, as the optical cavity is critically coupled to the coupling waveguide. Consequently, strong local heating at the location of the nanobeam only occurs when the laser is on resonance with the optical cavity mode and the intracavity photon number is maximized leading to increased absorption heating.

\subsection{Characterization of extended mechanical modes in the superfluid helium film}
\label{beamlike_mode_characterization}

\begin{figure*}
	\centering
	\includegraphics[width = 0.9\linewidth]{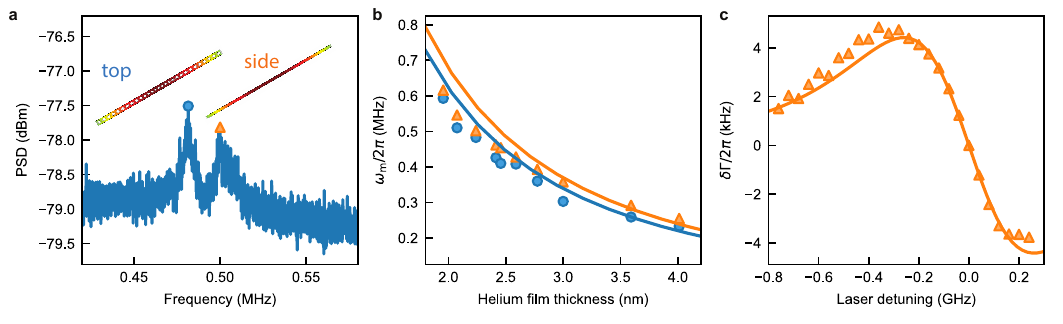}
	\caption{\textbf{Characterization of extended mechanical modes in the superfluid helium film.} \textbf{(a)} Power spectral density (PSD) of low-frequency helium mechanical modes measured through homodyne detection. \textbf{(b)} Dependence of mechanical mode frequency $\omega_\mathrm{m}$ on helium film thickness. Solid lines are results of finite-element simulations. \textbf{(c)} Change of linewidth $\delta \Gamma$ of the higher-frequency mechanical mode as a function of laser detuning from the optical cavity resonance. Solid line is a fit to the theory in Eq.~\eqref{delta_Gamma_final} including photothermal and radiation pressure coupling. Error bars for the data in \textbf{(b)} and \textbf{(c)} are not visible as they are smaller than the data point size.}
	\label{Fig:S3_beamlike_mechanical_modes}
\end{figure*}

In the low frequency regime of the homodyne spectrum (typically below frequencies of $\qty{1}{MHz}$), we observe a doublet of mechanical resonances of the superfluid helium film (see Fig.~\ref{Fig:S3_beamlike_mechanical_modes}(a)). Using finite-element simulations, we identify these resonances as extended third sound modes of the helium film. In particular, the lower mode in the doublet corresponds to a mode on the top and bottom surface of the nanobeam which is not affected by the phononic crystal patterning of the silicon material and is therefore extended across the whole beam. The upper mode in the doublet correponds to a mode on the sidewall of the nanobeam. As for the phononic crystal modes discussed in the main text, the mechanical frequency of these extended modes decreases strongly with increasing helium film thickness in good agreement with finite-element simulations (see Fig.~\ref{Fig:S3_beamlike_mechanical_modes}(b)) and the modes exhibit modifications of the mechanical linewidth due to (photothermal) backaction (see Fig.~\ref{Fig:S3_beamlike_mechanical_modes}(c)).

\subsection{Time-resolved detection of optomechanical coupling between helium mechanical and optical cavity modes}
\label{time_resolved_detection}

\begin{figure*}
	\centering
	\includegraphics[width = 0.95\linewidth]{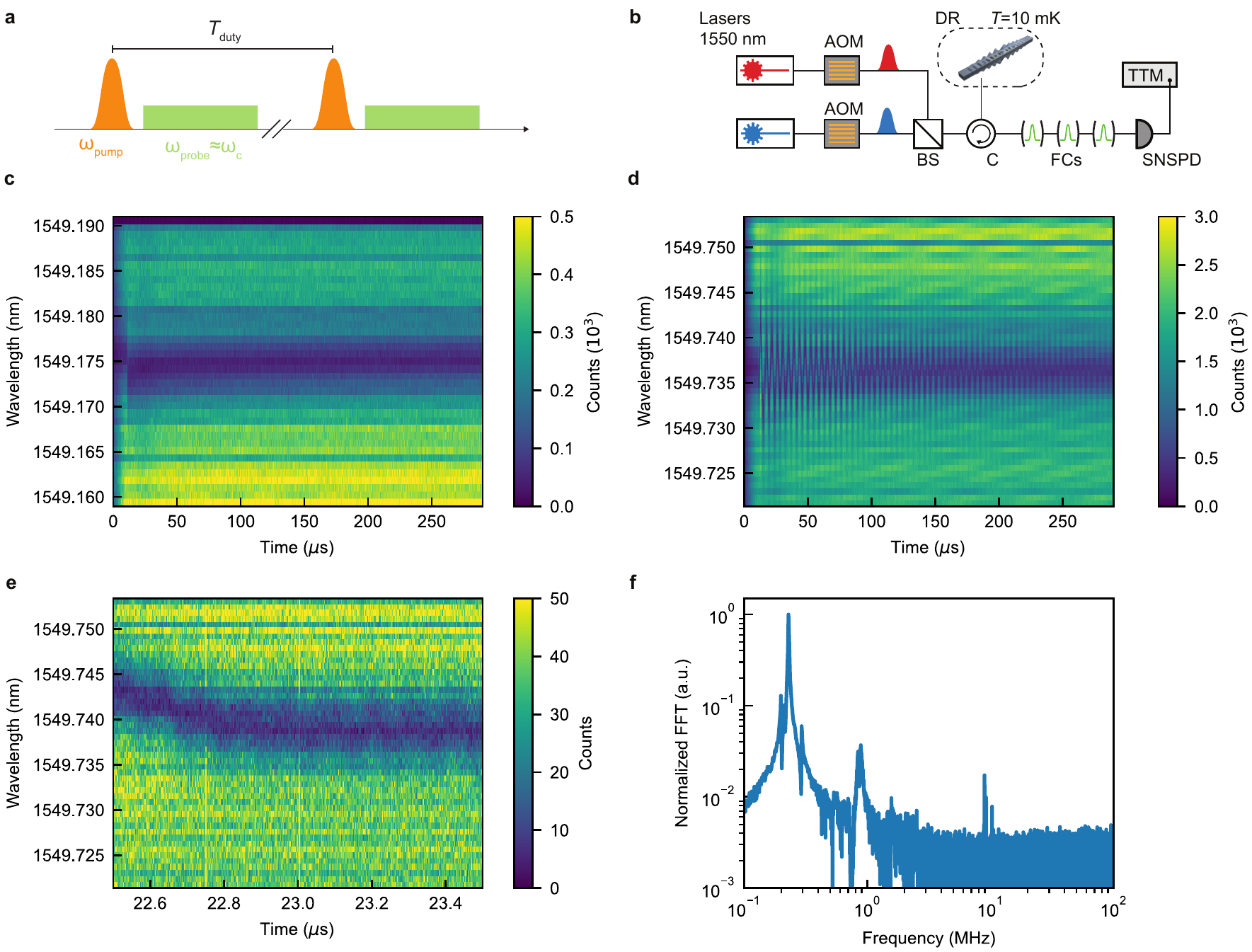}
	\caption{\textbf{Time-resolved detection of optical resonance shift.} \textbf{(a)} Pulse sequence for a pump-probe measurement of the optical resonance. A $\qty{100}{ns}$ pump pulse with power $P_\mathrm{pump}=\qty{100}{\micro\watt}$ at laser frequency $\omega_\mathrm{pump}$ off-resonant from the optical cavity is followed by a quasi-continuous weak probe pulse ($P_\mathrm{probe}=\qty{10}{fW}$). The frequency of the probe pulse $\omega_\mathrm{probe}$ is scanned across the optical cavity resonance at $\omega_\mathrm{c}$. The pulse sequence is repeated with a duty cycle of $T_\mathrm{duty}=\qty{1}{ms}$. \textbf{(b)} Measurement setup for pulsed measurements of the optical resonance. AOM, acousto-optic modulator; BS, beam splitter; C, circulator; FC, filter cavity; SNSPD, single-photon detector; TTM, time-tagging module; DR, dilution refrigerator. \textbf{(c)}, \textbf{(d)} Photon counts on the SNSPD as a function of time after the arrival of the pump pulse and at varying probe laser frequency without superfluid helium film \textbf{(c)} and with a $\qty{4}{nm}$ thick helium film \textbf{(d)}. \textbf{(e)} Zoom-in of \textbf{(d)}. \textbf{(f)} Normalized fast Fourier transform of the optical resonance wavelength as a function of time obtained from fits to data in \textbf{(d)}.}
	\label{Fig:S4_pulsed_mode_characterization}
\end{figure*}

In the main text, we use homodyne measurements to characterize third sound modes in the superfluid helium film. In addition, we directly resolve the time dynamics of the optical cavity induced by the presence of third sound modes by using a pulsed pump-probe measurement scheme (see Fig.~\ref{Fig:S4_pulsed_mode_characterization}(a)). A pump pulse at laser frequency $\omega_\mathrm{pump}$ induces motion of the superfluid helium modes. The pump pulse is detuned from the optical cavity resonance at $\omega_\mathrm{c}$ by $\Delta=\omega_\mathrm{pump}-\omega_\mathrm{c}=\qty{-3.5}{GHz}$. After the pump pulse, a quasi-continuous weak probe pulse is switched on to detect the optical cavity resonance.

The optical setup for pulsed excitation and readout of the helium mechanical motion is shown in Fig.~\ref{Fig:S4_pulsed_mode_characterization}(b). We create two optical pulses from two fiber-coupled lasers by using acousto-optic modulators. The pulses are combined on a fiber beam splitter and routed to the device inside the dilution refrigerator via a circulator. The reflected light from the ciculator is filtered by a series of three free-space Fabry-Perot cavities (linewidth $\qty{150}{MHz}$) locked at the frequency of the probe laser to remove the strong pump pulse. After filtering, the light is detected on a superconducting nanowire single-photon detector and the electronic signal is recorded by a time-tagging module~\cite{riedinger_remote_2018}.

We measure the photon counts detected on the single-photon detector as a function of time after the arrival of the pump pulse and repeat this measurement varying the frequency of the probe pulse; first without any superfluid helium in the chamber (Fig.~\ref{Fig:S4_pulsed_mode_characterization}(c)), and with a $\qty{4}{nm}$ superfluid helium film (Fig.~\ref{Fig:S4_pulsed_mode_characterization}(d)). Without superfluid helium, a dip in the reflection spectrum is observed at the center wavelength of the optical resonance at $\lambda_\mathrm{c}=\qty{1549.175}{nm}$. The center wavelength remains constant as a function of time except for a small blue shift at time $t=\qty{10}{\micro s}$ when the pump pulse arrives at the device. The blue shift of the optical cavity resonance is likely related to free-carrier dispersion effects from the strong pump pulse which are commonly observed in silicon photonic crystal cavities~\cite{barclay_nonlinear_2005,chan_laser_2011}. When the nanobeam is covered with a $\qty{4}{nm}$ superfluid helium film, we observe strong oscillations of the optical resonance frequency after arrival of the pump pulse with an amplitude larger than the optical cavity linewidth (see Fig.~\ref{Fig:S4_pulsed_mode_characterization}(d)) and period of $\qty{5}{\micro s}$. In addition, oscillations on faster time scales with a period of approximately $\qty{120}{ns}$ and smaller amplitude are observed (see Fig.~\ref{Fig:S4_pulsed_mode_characterization}(e)).

By fitting the optical resonance in Fig.~\ref{Fig:S4_pulsed_mode_characterization}(d) at each point in time, we obtain the spectrum of the oscillations through Fourier transformation as shown in Fig.~\ref{Fig:S4_pulsed_mode_characterization}(f). We observe dominant peaks in the spectrum at frequencies of $\qty{200}{kHz}$ and around $\qty{10}{MHz}$, which correspond to the slow and fast time scale oscillations observed in Figs.~\ref{Fig:S4_pulsed_mode_characterization}(d) and (e). By comparing the data to the results of finite-element simulations of third sound modes in the $\qty{4}{nm}$ superfluid helium film, we identify the peaks at $\qty{200}{kHz}$ with extended third sound modes discussed in Sec.~\ref{beamlike_mode_characterization} and the peaks at $\qty{10}{MHz}$ with the confined phononic crystal modes discussed in the main text. An additional broad peak at a frequency of $\qty{1}{MHz}$ appears in the spectrum in Fig.~\ref{Fig:S4_pulsed_mode_characterization}(f), which is not observed in homodyne measurements. While the origin of this peak cannot be clearly identified based on current measurements, it is likely related to a weakly confined higher-order extended third sound mode of the superfluid film.

\section{Impact of superfluid helium on 5-GHz mechanical breathing mode of the nanobeam resonator}
\label{impact_helium_breathing_mode}

\begin{figure}
	\centering
	\includegraphics[width = 0.5\linewidth]{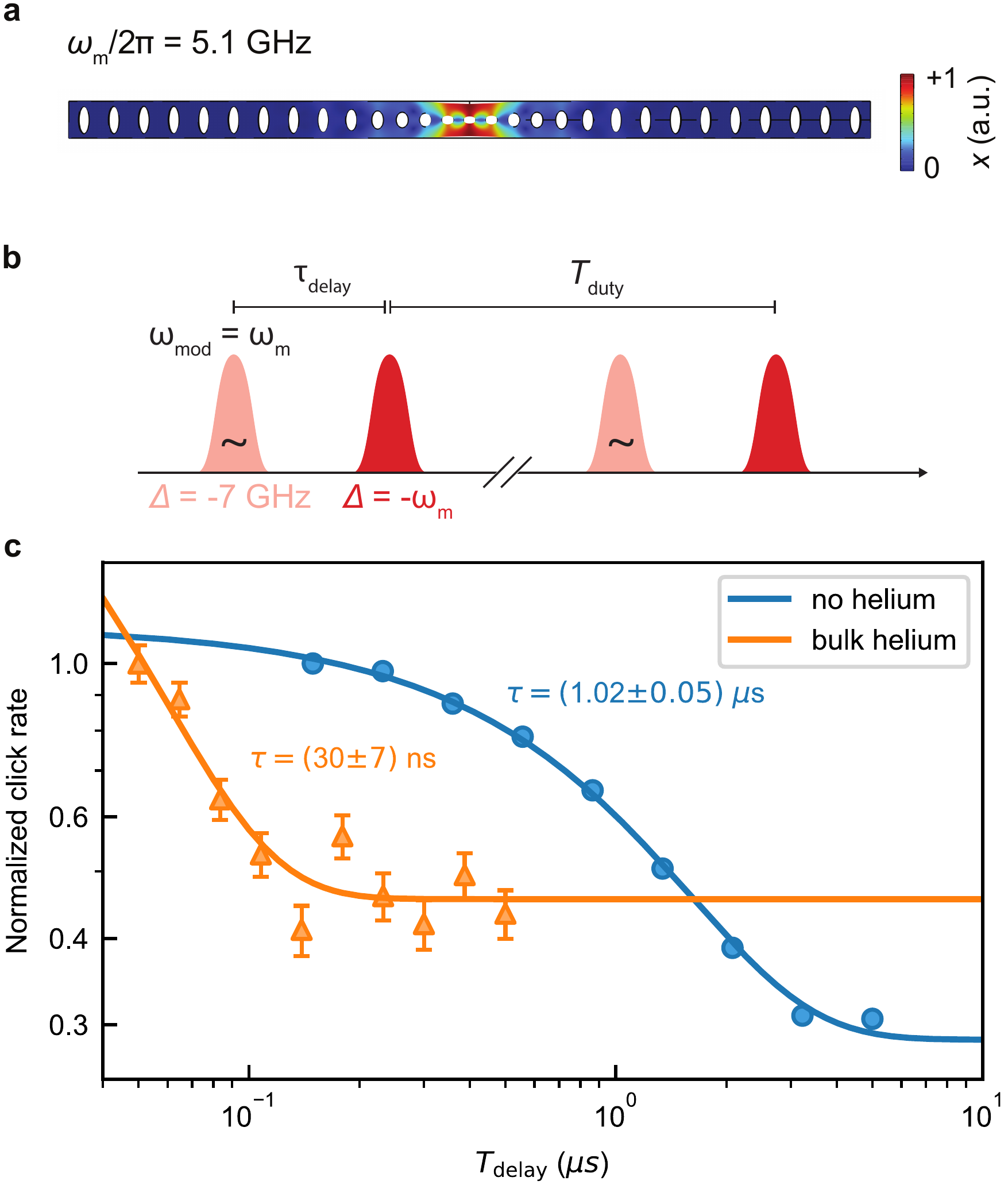}
	\caption{\textbf{Lifetime measurement of breathing mode of the silicon nanobeam.} \textbf{(a)} Simulated displacement mode shape of silicon mechanical breathing mode at frequency $\omega_\mathrm{m}/2\pi=\qty{5.1}{GHz}$. \textbf{(b)} Pulse sequence used for mechanical lifetime measurement. First, a pulse detuned from the optical cavity by $\Delta=-7$ GHz and modulated by an electro-optic phase modulator at the mechanical frequency $\omega_\mathrm{m}$ coherently excites a large phonon population in the breathing mode of the nanobeam. After a delay of $\tau_\mathrm{delay}$, a pulse detuned at the red sideband of the optical cavity resonance $\Delta=-\omega_\mathrm{m}$ reads out the phonon population through the optomechanical anti-Stokes scattering process. The pulse sequence is repeated with a duty cycle of $T_\mathrm{duty}=10~\mu$s \textbf{(c)} Count rates during the read-out pulse without helium and with the nanobeam immersed in bulk superfluid helium as a function of the pulse delay $\tau_\mathrm{delay}$. The count rates are normalized to the first data point of each data set. Solid lines correspond to a fit to an exponential decay with lifetime $\tau$. Error bars for the data without helium are not visible as they are smaller than the data point size.}
	\label{Fig:S6_coherent_drive}
\end{figure}

The nanobeam resonators used in this work also support mechanical modes in the silicon material itself~\cite{chan_optimized_2012}. In particular, the fundamental breathing mode at a frequency of $\omega_\mathrm{m}/2\pi =\qty{5.1}{GHz}$ (see Fig.~\ref{Fig:S6_coherent_drive}(a)) is confined to the defect region of the nanobeam and couples to the optical cavity resonance through a combination of photoelastic coupling and moving boundary conditions \cite{chan_optimized_2012}. This optomechanical interaction has been used in a variety of experiments with mechanical resonators in the quantum regime making silicon nanobeams a prototypical system for quantum optomechanics \cite{chan_laser_2011,riedinger_non-classical_2016,hong_hanbury_2017,riedinger_remote_2018,maccabe_nano-acoustic_2020,wallucks_quantum_2020}. Hence, we study the effect of superfluid helium on the silicon mechanical breathing mode.

\subsubsection{Mechanical damping in bulk superfluid}
\label{mechanical_damping_in_bulk_superfluid}

In contrast to the previous experiments, we now fill the helium chamber with additional helium gas until the optical resonance does not shift any further. At this point the nanobeam is immersed in a bulk superfluid environment. We study the damping of the nanobeam breathing mode due to the additional coupling to the bulk superfluid helium, using a pulsed measurement of the phonon lifetime. The pulse sequence is show in Fig.~\ref{Fig:S6_coherent_drive}(b). First, we send a pump pulse which is detuned from the optical cavity resonance by $\Delta=\qty{-7}{GHz}$ and thus not resonant with the optomechanical sideband. This pulse is phase modulated by an electro-optic modulator at the mechanical frequency $\omega_\mathrm{m}/2\pi = \qty{5.1}{GHz}$ to coherently drive the mechanical mode. Phonons are read out by a second pulse detuned on the red optomechanical sideband $\Delta=-\omega_\mathrm{m}$ at a pulse delay $\tau_\mathrm{delay}$.

Figure~\ref{Fig:S6_coherent_drive}(c) shows the click rate of optomechanically scattered photons from the read-out pulse as a function of pulse delay $\tau_\mathrm{delay}$. The phonon lifetime $\tau$ obtained from an exponential fit to the data points decreases from $\tau=\qty{1}{\micro s}$ when there is no superfluid helium in the chamber to $\tau=\qty{30}{ns}$ when the same device is immersed in bulk superfluid helium. The measured lifetime is most likely limited by the length of the optical pulses used, and the real value may be even lower. We attribute this significant decrease in phonon lifetime to damping of the mechanical motion due to the superfluid helium environment. Even though there is no viscous damping in superfluid helium, the motion of the nanobeam sidewalls can transfer energy to the superfluid and therefore emit phonons into a continuum of phonon modes. This mechanism has been shown to lead to strong damping for mechanical modes in nanostrings at $\omega_\mathrm{m}/2\pi \approx \qty{1}{MHz}$ \cite{fong_phonon_2019} and whispering gallery microdisk resonators at mechanical frequency $\omega_\mathrm{m}/2\pi \approx \qty{800}{MHz}$ \cite{guenault_probing_2019}.

\subsection{Optomechanical thermometry in presence of superfluid thin-film}

We further perform sideband thermometry measurements in the presence of superfluid thin films on the silicon nanobeam to determine the phonon occupation of the mechanical mode and study the impact of the superfluid helium. The asymmetry of optomechanical photon scattering rates from the blue or red sideband can be used to calibrate the mechanical phonon occupancy. However, as this measurement depends crucially on the exact laser detuning with respect to the optical cavity resonance, oscillations of the optical resonance frequency due to third sound waves as discussed in Sec.~\ref{time_resolved_detection} inhibit obtaining reliable results from these measurements. Notably, from these measurement we do not find any indication that the presence of a superfluid thin film reduces the thermal occupation as might be speculated due to improved thermal anchoring of the mechanical breathing mode to the cryogenic environment through the high thermal conductivity of helium.

\end{document}